\documentclass[12pt]{article}
\usepackage{tikz}
\usetikzlibrary{decorations.pathreplacing,angles,quotes}
\usepackage{etex}
\usepackage[T1]{fontenc}
\usepackage{amssymb,amsmath,geometry}
\usepackage{mathtools}
\usepackage{graphicx}
\usepackage{fancyhdr}
\usepackage{fancybox}
\usepackage{shadow}
\usepackage{empheq}
\usepackage{titlesec}
\usepackage{titletoc}
\usepackage{soul, color}
\usepackage{float}
\usepackage{shorttoc}
\usepackage{xcolor}
\usepackage{fancybox}
\usepackage{natbib}
\usepackage{array}
\usepackage{tabularx}
\usepackage{booktabs}
\usepackage{rotating}
\usepackage{bigstrut}
\usepackage{natbib}
\usepackage{aeguill}
\usepackage{changepage}
\newenvironment{changemargin}[2]{\begin{list}{}{%
\setlength{\topsep}{0pt}%
\setlength{\leftmargin}{0pt}%
\setlength{\rightmargin}{0pt}%
\setlength{\listparindent}{\parindent}%
\setlength{\itemindent}{\parindent}%
\setlength{\parsep}{0pt plus 1pt}%
\addtolength{\leftmargin}{#1}%
\addtolength{\rightmargin}{#2}%
}\item }{\end{list}}

\usepackage{hyperref}
\def\sym#1{\ifmmode^{#1}\else\(^{#1}\)\fi}
\usepackage{caption}
\usepackage{blindtext}

\usepackage{caption}
\usepackage{subcaption}
\usepackage{mathtools}

\usepackage{graphicx}

\usepackage{enumerate}

\usepackage{hyperref}
\hypersetup{backref,
colorlinks=true, citecolor=blue!50!black}
\usepackage{graphicx}

\renewcommand{\thetable}{\Roman{table}}
\renewcommand{\thefigure}{\Roman{figure}}

\usepackage{titlesec}
\usepackage{setspace}
\usepackage{indentfirst}

\makeatletter
  \newcommand\smalls{\@setfontsize\smalls{10.3pt}{6}}
\makeatother

\makeatletter
  \newcommand\footnotesizes{\@setfontsize\footnotesizes{9.6pt}{6}}
\makeatother

\newsavebox\tmpbox

\usepackage{caption}
\usepackage{subcaption}

\title{Armed Conflict and Early Human Capital Accumulation:  Evidence from Cameroon's \\Anglophone Conflict}
\author{Hector Galindo-Silva\thanks{%
Department of Economics, Pontificia Universidad Javeriana, E-mail: galindoh@javeriana.edu.co
} \\
Pontificia Universidad Javeriana\\
 \and Guy Tchuente \thanks{%
School of Economics, University of Kent, E-mail: guytchuente@gmail.com} \\
University of Kent
}
\begin{document}
\maketitle
\begin{abstract}This paper examines the impact of the Anglophone Conflict in Cameroon on human capital accumulation. Using high-quality individual-level data on test scores and information on conflict-related violent events, a difference-in-differences design is employed to estimate the conflict's causal effects. The results show that an increase in violent events and conflict-related deaths  causes a significant decline in test scores in reading and mathematics. The conflict also leads to higher rates of teacher absenteeism and reduced access to electricity in schools. These findings highlight the adverse consequences of conflict-related violence on human capital accumulation, particularly within the Anglophone subsystem. The study emphasizes the disproportionate burden faced by Anglophone pupils due to language-rooted tensions and segregated educational systems. \\%The research contributes to the understanding of the effects of language-based conflicts on human capital and provides valuable insights for post-conflict reconstruction efforts in the education sector.\\
\end{abstract}

\newpage

%%%%%%%%%%%%%%%%%%%%%%%%%%%%%%%%%%%%%%%%%%%%%%%%%%%%%%%%%%%%%%%%%%%%%%%%%%%%
\section{Introduction}

Since 2016, Cameroon, a central African country, has been immersed in a violent civil conflict between armed separatist groups and the government. Referred to as the Anglophone Conflict or the Anglophone Crisis, this conflict has been primarily concentrated in two regions where English is the predominant language, encompassing approximately 20\% of the country's landmass. Its consequences have been devastating, resulting in a significant loss of life, with reported casualties exceeding 4,000, and causing the displacement of more than 788,000 individuals, leading to widespread population movements \citep{HRWworlreport22}.

The Anglophone Conflict in Cameroon is primarily driven by a linguistic division rooted in the country's colonial history. With English and French as official languages,\footnote{It is worth noting that although Cameroon boasts over 200 local languages, French and English serve as the official languages, aiming to unify its ethnically diverse population.} the Francophone population holds power in the government and elite circles, while some Anglophones have long faced marginalization. The conflict, which began in 2016 as a peaceful protest by English-speaking lawyers and teachers, has since escalated into a violent clash between armed separatists and the Cameroonian government.\footnote{Numerous reports indicate that English-speaking separatists, advocating for the establishment of an independent English-speaking state called Ambazonia, have terrorized civilians and engaged in attacks against government forces \citep{HRWnews21}. Additionally, there are reports of Cameroonian troops firing upon unarmed civilians and destroying their homes \citep{HRWnews18, Amnesty18}.} 

A significant aspect of the violence associated with this conflict emerged in 2017 when Anglophone activists initiated `Operation Ghost Town.' This protest called for the closure of Anglophone schools in the north-west and south-west regions to oppose the perceived assimilation of the Anglophone education system into the French-speaking one.\footnote{As explained later in this paper, the education system in Cameroon is divided into Anglophone and Francophone subsystems. While Anglophone schools are primarily located in certain regions of the country, there is geographic overlap with Francophone schools. Nevertheless, despite this overlap, the call for school closures by Anglophone activists specifically targeted Anglophone schools.} The boycott has been marked by violence, including student abductions and targeted killings of school staff \citep{HRWnews18, HRWnews21, Amnesty18, OCHAreport2021}. These incidents have become critical flashpoints in the ongoing conflict \citep{TheGuardian092018, TheWashingtonPost092018}.

This paper investigates the impact of violence associated with Cameroon's Anglophone Conflict on the acquisition of human capital by pupils. The unique characteristics of this conflict, such as its localized geography and linguistic grievances as the primary driver, present an exceptional opportunity to analyze the consequences of armed conflicts associated with language-related matters. Moreover, the coexistence of both Anglophone and Francophone education systems in Cameroon, with overlapping geographic areas, provides a natural laboratory to explore the implications for human capital resulting from conflicts of this nature.

Our analysis utilizes two main data sources: pupil test scores in reading and mathematics for Grades 2 and 6, obtained from the Programme d'Analyse des Systemes Educatifs de la CONFEMEN (PASEC) for the years 2014 and 2019. This data includes a representative sample of pupils from both the Francophone and Anglophone subsystems, allowing for a comparison before and after the onset of the Anglophone Conflict. Additionally, we rely on the Armed Conflict Location \& Event Data Project (ACLED) to gather information on violent events and fatalities associated with the conflict. Spanning from 2000 to 2022, this dataset provides comprehensive information including the date, location, and actors involved in each event (such as Ambazonian rebels or the Cameroonian army).

 Our study focuses on pupils in the conflict-affected North-West and South-West regions who are enrolled in the Anglophone subsystem.  We compare this group, referred to as the treatment group, with Anglophone subsystem pupils in unaffected regions as the control group. Using a difference-in-differences methodology, we analyze the causal impact of the Anglophone Conflict on human capital accumulation. Our primary identification assumption is that, in the absence of the conflict and given certain observable variables (included in our analysis), the change in test scores between 2014 and 2019  would have been the same for all pupils.  We provide evidence consistent with this assumption using data from  pupils from the francophone subsystem during the same time period. 
 
Our main finding reveals a significant and negative causal relationship between violence stemming from the conflict and the accumulation of human capital among pupils in the North-West and South-West regions of the Anglophone educational subsystem. Specifically, our estimates indicate that an additional ten violent events involving Ambazonian rebels correspond to a 2.6\% decrease in reading test scores and a 2.1\% decline in mathematics test scores for students enrolled in the Anglophone educational subsystem. Furthermore, for every ten deaths, there is a 1.9\% decrease in language test scores and a 1.9\% reduction in mathematics test scores for these pupils. These results indicate that events resulting in loss of life have a detrimental impact on the accumulation of human capital. Given the crucial role of early childhood years in human capital development, these declines may have significant long-term consequences (which we cannot directly study due to the temporal limitation of our data).

Moreover, our analysis indicates that a rise in fatal events involving Ambazonian rebels is linked to an increased rate of teacher absenteeism in the Anglophone education subsystem. This finding implies that the presence of a high-risk environment detrimentally impacts both the quality and quantity of pupils' learning experiences, ultimately leading to a diminished acquisition of skills. Additionally, our research demonstrates a negative association between conflict-related fatalities and schools' access to electricity. This observation implies that the presence of conflict exacerbates economic hardships, potentially leading pupils to shift their focus from studying to engaging in labor within either the legal or illegal economy. As a consequence, this further widens the disparities in learning outcomes among Anglophone pupils.

We also investigate the effect of the Anglophone conflict on educational outcomes of pupils enrolled in the Francophone education subsystem (in the Grand-Ouest and Grand-Nord regions). Due to the characteristics of the Anglophone conflict, we anticipate that these students have been less directly exposed to armed violence or have experienced it indirectly. Our analysis reveals that fatal events involving Ambazonian rebels have considerably smaller effects on their human capital accumulation compared to their Anglophone counterparts. In fact, the observed effects are approximately 10 times smaller for these pupils. 

Our research not only highlights the detrimental impact of violence related to the Anglophone conflict on the accumulation of human capital but also sheds light on a mechanism by which ethnic or linguistic disparities are exacerbated within the context of armed conflict. Remarkably, this conflict either originates from or is driven by the intent to reduce these differences. Our study demonstrates that such exacerbation occurs not only in the short term, where there is direct destruction of lives and infrastructure, particularly in regions with a predominantly English-speaking population, but also in the long term, as early-life experiences significantly shape individuals' future educational and economic trajectories \citep{heckman2013understanding}.  Specifically, we identify a discernible effect of the Anglophone conflict on students within the Anglophone and Francophone educational subsystems.

The existing body of literature on the impact of armed conflict on human capital is extensive.  Previous studies, such as  \cite{bertoni2019education} , have demonstrated the significant consequences of armed conflicts on the accumulation of human capital. Furthermore, exposure to violence during childhood has been linked to negative effects on various aspects of individuals' lives, including psychological well-being, health, and education, as indicated by \cite{jurges2022cohort}.

Numerous studies have provided evidence of the detrimental effects of conflicts on school attainment, academic achievement, and educational performance. These studies, including those by \cite{akbulut2014children, bertoni2019education, chamarbagwala2011human, dabalen2014estimating, di2013effect, leon2012civil, pivovarova2015quantifying, singh2016gender, swee2015war, verwimp2014schooling, valente2014education, bruck2019learning}, have significantly contributed to our understanding of the relationship between conflict and human capital accumulation—the acquisition of skills, knowledge, and experience through formal education that are valuable in the production process.

However, the specific impacts of language-based violence remain poorly understood. Existing research has only tangentially explored how ethnic disparities in relate these effects, with limited attention paid to this aspect in studies by \cite{chamarbagwala2011human, dabalen2014estimating, bertoni2019education}. Importantly, none of these studies have specifically focused on conflicts rooted in language identity, which constitutes the primary focus of this paper.

Moreover, our study makes a significant contribution to this literature by providing causal estimates of the effects of language-rooted conflicts. We demonstrate that when educational systems are segregated,the minority group disproportionately bears the burden of the conflict. To the best of our knowledge, this paper is the first to propose a quasi-experimental design using a difference-in-differences approach, complemented by high-quality measures of individual skill levels, to examine the effects of the Anglophone conflict in Cameroon. Consequently, our research contributes to the ongoing interdisciplinary efforts aimed at assessing the consequences of the Anglophone crisis and exploring potential peaceful resolutions \citep{willis2023moral, willis2019human, nwati2020anglophone, crawford2021shrinking, pelican2022anglophone, ousmanou2022internally, pelican2022mbororo}.

The remainder of the paper proceeds as follows. Section 2 presents details regarding the data sources and institutional background pertinent to early human capital and the Anglophone Conflict. Section 3 describes our research design. Section 4 examines the impact of armed conflict on early human capital accumulation. Section 5 focuses on estimating the effects of the conflict on the Francophone subsystem. Finally, Section 6 concludes the paper.

%%%%%%%%%%%%%%%%%%%%%%%%%%%%%%%%%%%%%%%%%%%%%%%%%%%%%%%%%%%%%%%%%%%%%%%%%%%%

\section{Institutional Background and Data Sources}

\subsection{ Cameroon's Primary Schools Education System}
Cameroon's primary education sector is served by three main providers: the government, responsible for public schools, private institutions, and private confessional schools such as Catholic, Islamic, and Protestant establishments. These providers deliver education in the country's two official languages: French and English, each with its own separate primary education subsystem.

The primary education system in Cameroon is characterized by the presence of both French and English subsystems in all ten regions of the country. However, the majority of regions (eight out of ten) are predominantly French-speaking, resulting in a higher concentration of French schools. On the other hand, English-speaking regions like the North-West and South-West have a greater proportion of educational institutions that offer instruction in the English language.

Data collected during the 2014-2015 academic year shows that 71.6\% of pupils were enrolled in the Francophone education system, while 28.4\% were enrolled in the Anglophone education system \citep{alemnge_curriculum_2019}. It is worth noting that 75\% of pupils attend public schools, and there is a higher proportion of pupils in the Francophone system compared to the Anglophone system \citep{alemnge_curriculum_2019}.

\begin{figure}[!h] 
\centering
\caption{Enrolment Rate 1994 to 2019 from the World Bank data.}
\label{Enroltrend2}
\includegraphics[width=0.7\linewidth]{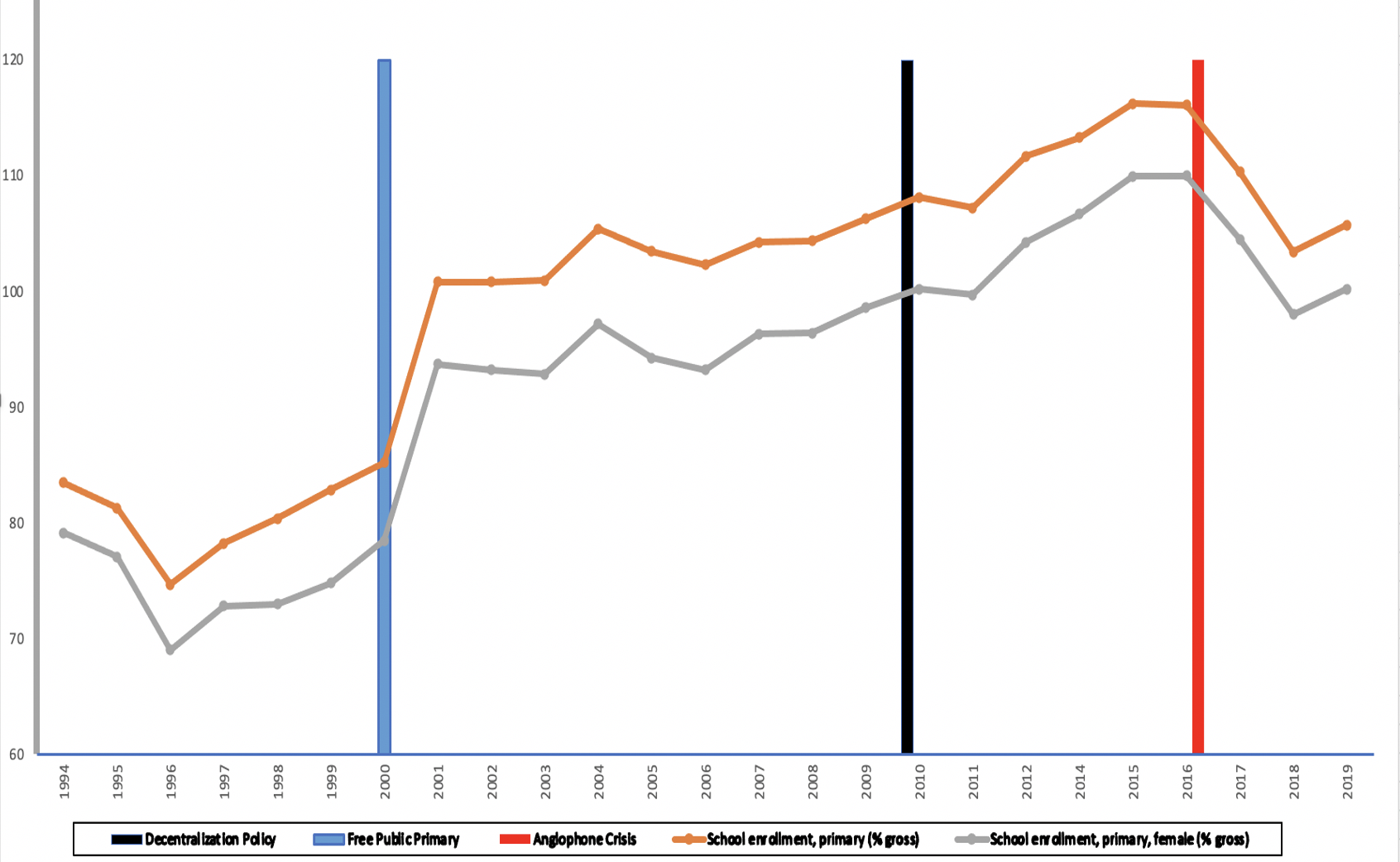}
\end{figure}

Regarding the evolution of the educational system, Figure \ref{Enroltrend2} illustrates a consistent upward trend in primary school enrollment rates for both boys and girls from 1996 to 2016. Notably, there were periods of accelerated growth in enrollment rates in 2000 and 2010. However, in 2016, there was a decline in enrollment rates for both genders. This decline may be attributed to the ongoing conflict in the Anglophone regions since 2016. While this graph indicates a reduction in the percentage of primary school enrollment around the onset of the Anglophone crisis, the precise causal impact of the conflict on human capital accumulation remains to be accurately determined. Moreover, further analysis is necessary to gain a comprehensive understanding of the effects of the conflict on primary school enrollment rates. This paper will focus on its impact on the accumulation of human capital.

\subsection{Cameroon's Anglophone Conflict}

\begin{figure}[H]
             \caption{Conflict-Related Events in Cameroon}
        \label{fig:ACLED_eventfatalities}
\begin{subfigure}{0.5\textwidth}
\caption{Any group involved} \label{fig:ACLED_eventfatalitiesby}
\includegraphics[width=\linewidth]{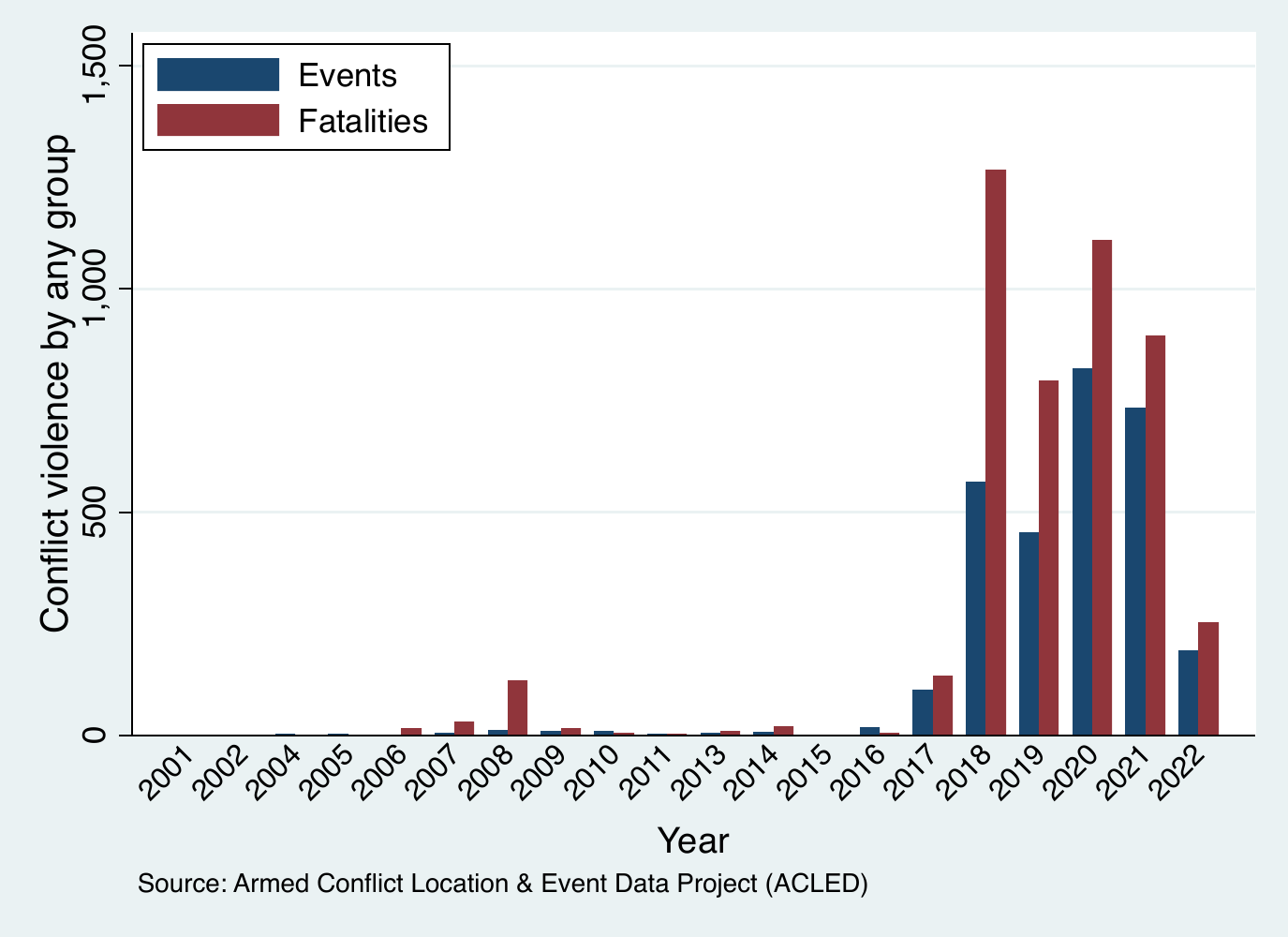}
\end{subfigure}\hspace*{\fill}
\begin{subfigure}{0.5\textwidth}
\caption{Ambazonian Separatists involved} \label{fig:Envent_conflict_AMB}
\includegraphics[width=\linewidth]{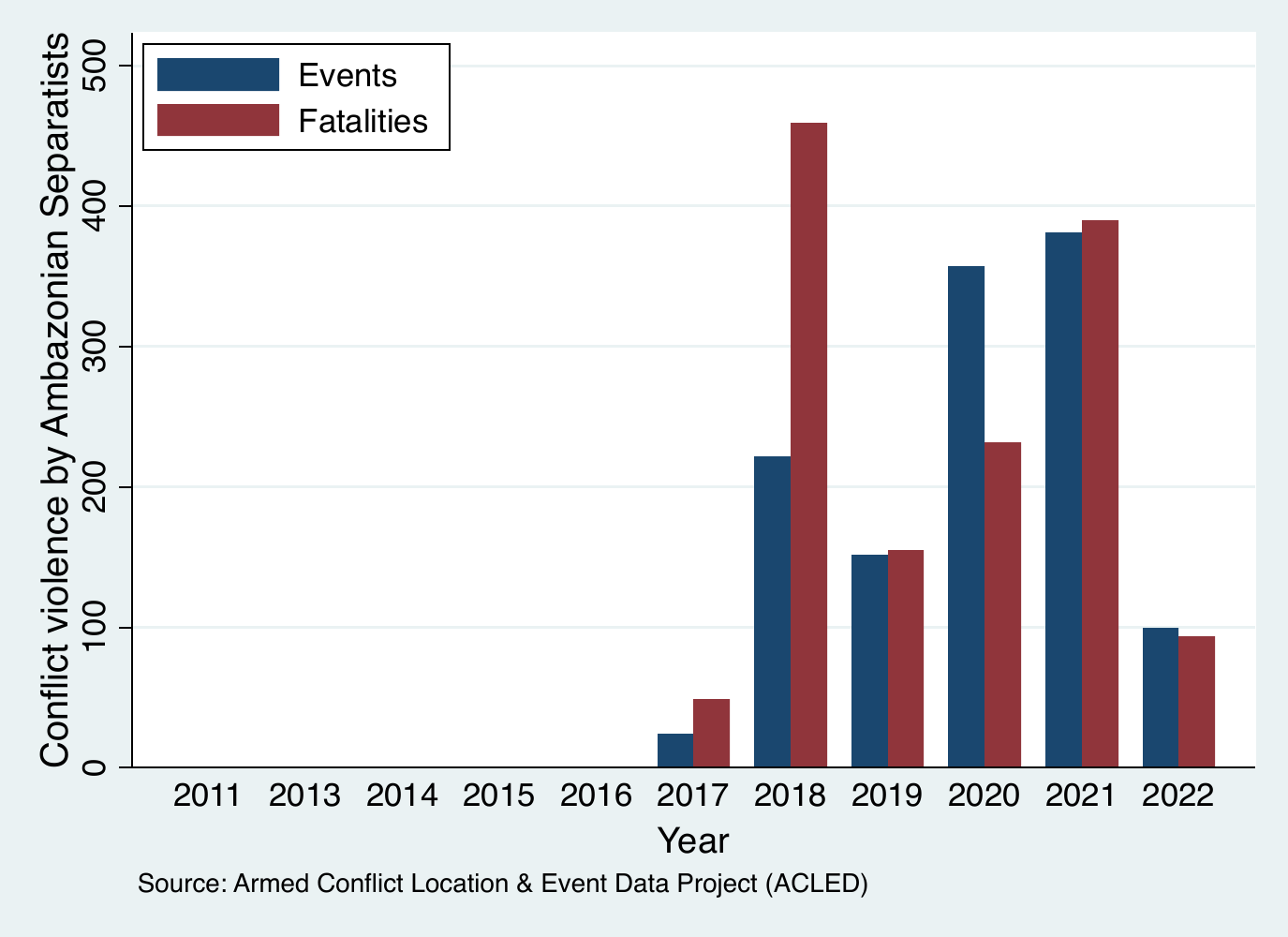}
\end{subfigure}
\end{figure}

Cameroon gained independence from French and English colonial rule in 1960 and 1961, respectively, for the Francophone and Anglophone regions, and has generally enjoyed a peaceful environment without military coups. However, recent years have seen the emergence of conflicts in Cameroon that pose threats to its stability. To illustrate this, Figure \ref{fig:ACLED_eventfatalitiesby} provides a visual representation of conflict-related events and fatalities since 2001.

Among the conflicts experienced by Cameroon in recent years, the Anglophone Conflict has emerged as one of the most intense.\footnote{In addition to the Anglophone conflict, Cameroon has faced the insurgency of Boko Haram in the Far North, as well as complex challenges related to refugees and transborder issues in the East region. \citep[see][for a detailed background and literature review on the Anglophone conflict]{pelican2022anglophone}.}   It primarily stems from a linguistic division, with its epicenter located in the North West and South West regions.  The main actors involved are the Cameroonian army and the Ambazonian separatists. The origins of the Anglophone Conflict can be traced back to the historical legacy of Cameroon's two official languages, English and French, inherited from the colonial era. The institutional framework established by former colonial powers, resulted in the dominance of French speakers in Cameroon's government and elite circles, and contributed to the marginalization of the Anglophone population. This has given rise to various social issues commonly referred to as the `Anglophone problem' \citep{konings1997anglophone}. 

In 2016, the `Anglophone problem' transformed into a full-fledged violent conflict. This escalation originated from peaceful protests organized by English-speaking lawyers and teachers. They were driven by their frustrations with the government's practice of assigning French-speaking judges and teachers to English-speaking courts and schools. The English-speaking community argued that this forced assimilation into Francophone legal and educational systems. While the government initially acknowledged the need for some reforms, it simultaneously repressed activists by imprisoning moderate leaders and employing violence against protesters. As moderate voices were silenced, more extremist factions emerged, advocating for complete separation from Cameroon and demanding independence \citep{TheWashingtonPost092018, pelican2022anglophone}. Subsequently, the conflict intensified, with separatist groups increasing their attacks on the military, prompting retaliatory actions by Cameroonian troops. This retaliation has included firing upon unarmed civilians and demolishing their homes. 

One notable series of events in this conflict is known as the `Operation Ghost Town,' which was initiated by Anglophone activists in 2017. This `operation' aimed to achieve the closure of schools in the north-west and south-west regions as a protest against the perceived assimilation of the Anglophone education system into the French-speaking one. Notably, this `operation' specifically targeted Anglophone schools in these regions, while excluding Francophone schools operating in the same areas. Acts of violence associated with this `operation' have included forced closures of schools, the abduction of students, and targeted killings of principals and staff members within these Anglophone schools.

In addition to including targeted violence against Anglophone schools, the Anglophone conflict is characterized by two central features. Firstly, it unfolds within a relatively concise timeframe, with violence stemming from this conflict first appearing in 2017 and rapidly escalating in 2018. This pattern is evident in Figure \ref{fig:Envent_conflict_AMB}, which illustrates the distribution of conflict-related events and fatalities specifically attributed to Ambazonian separatists from 2011 to 2022.\footnote{Figure \ref{fig:Envent_conflict_AMB} also highlights the escalation of the conflict in 2020 and 2021, leading to a substantial rise in both fatalities and events. Despite the Major National Dialogue held in October 2019, which engaged multiple stakeholders, the desired political transformation and decrease in violence within the Anglophone regions were not realized, as depicted in Figure \ref{fig:ACLED_eventfatalities}.} 

The second notable characteristic of the Anglophone conflict is its geographical localization, confined exclusively to the North West and South West regions of Cameroon. This remarkable geographic limitation is visually depicted in Map \ref{Fatal_AMB_map} of the Appendix. Given the significance of this aspect to our empirical analysis, we will provide a more comprehensive exploration in Section \ref{Distribution_of_Conflict_Related_Events}.

\subsection{Data sources}

The objective of this study is to examine the effect of violence associated with the Anglophone conflict on the academic achievements of pupils in different regions of Cameroon. In order to evaluate this impact, we utilize reading and mathematics test scores from the Programme d'Analyse des Systemes Educatifs de la CONFEMEN (PASEC) dataset. This dataset, collected in 2014 and 2019, comprises a representative sample of pupils enrolled in schools across French-speaking African nations and can be accessed publicly through registration on the official PASEC website.\footnote{The PASEC dataset can be accessed at \url{https://pasec.confemen.org/en/}}

The PASEC data on Cameroon consists of information from approximately 180 schools and around 4,887 pupils. This dataset encompasses both the francophone and anglophone education subsystems and includes data from all ten regions of the country.  The PASEC data is obtained from a sample of individuals organized into `strates', which represent the regions and educational subsystem.  In 2014, there were six strates, but in 2019, this number increased to twelve.\footnote{For a comprehensive list of these strates, please refer to Table \ref{strates1419} in Appendix \ref{Educationdata}.}  To ensure comparability between the two years, as explained in detail in Appendix \ref{Educationdata}, we focus on the 2014 strates. Table \ref{tab:stratesmain} presents these strates along with their corresponding regions and educational subsystems.

\begin{table}[H]
\centering 
{\renewcommand{\arraystretch}{0.9}
\setlength{\tabcolsep}{10pt}
\caption{Strates along with their corresponding regions and educational subsystems}
\footnotesize
  \begin{tabular}{lp{70mm}l}
\hline \addlinespace[0.1cm]
   \multicolumn{1}{c}{\textbf{Strate}} & \multicolumn{1}{c}{\textbf{Region}}  & \multicolumn{1}{c}{\textbf{Educational system}}  \\\addlinespace[0.1cm]\hline\addlinespace[0.1cm]
1 (Zone Anglophone)  & North-West,  South-West & Anglophone (public)  \\\addlinespace[0.1cm]
 2  (Zone Anglophone) &   North-West,  South-West & Anglophone  (private) \\\addlinespace[0.1cm]
3  (Zone Francophone)  & Littoral, West, Centre, Sud, Est,  \newline Adamaoua, North, Extrem-Nort  & Anglophone \\\addlinespace[0.1cm]
  4  (Grand-Ouest) &  West, North-West,  South-West, Littoral & Francophone \\\addlinespace[0.1cm]
  5 (Grand-Centre) &  Centre, Sud, Est & Francophone \\\addlinespace[0.1cm]
 6 (Grand-Nord) &  Adamaoua, North, Extrem-Nort  & Francophone \\
\addlinespace[0.1cm]\hline\addlinespace[0.1cm]
\end{tabular}
\label{tab:stratesmain}
}
\end{table}

As shown in Table \ref{tab:stratesmain}, strates 1 and 2 (referred to as Zone Anglophone) encompass pupils from both public and private schools within the Anglophone subsystem residing in the North-West and South-West regions of Cameroon (the English-speaking regions). As previously mentioned, these regions are known to be the theatre of the Anglophone conflict. Strate 3 (or Zone Francophone) comprises pupils from the Anglophone subsystem attending public schools in all regions except the North-West and South-West (i.e., the French-speaking regions of Cameroon). Strate 4 (or Grand-Ouest), includes pupils from the Francophone subsystem from the West, North-West,  South-West, and Littoral regions.  It is worth noting that Strate 4 overlaps geographically with Strate 1. Additionally,  Strate 5 (or Grand-Centre) encompasses pupils selected from the Centre, Sud, and Est regions. Lastly, Strate 6 (or Grand-Nord) comprises pupils from the Adamaoua, North, and Extrem-North regions.\footnote{It should be noted that strates 1 and 4 overlap in terms of the geographical regions they cover.}

Our study focuses on the Cameroonian regions that have been directly impacted by violence associated with the Anglophone conflict. Specifically, we aim to analyze the effects of this violence on Anglophone public schools. Given this objective, the information pertaining to strates 1 and 3 holds particular relevance for our research. In the subsequent discussion, we will elaborate on the definition of strate 1 as our control group, while strate 3 will serve as our treatment group. Descriptive statistics for our primary academic outcome variables are presented separately for these two strates, as well as by year, in Panels A and C of Table \ref{precharacteristics_tab}.

 \begin{table}[H]
 \renewcommand{\arraystretch}{0.7}
\setlength{\tabcolsep}{3pt}
\begin{center}
\caption {Descriptive Statistics}  \label{precharacteristics_tab}
\vspace{-0.2cm}
\footnotesize	
 \begin{tabular}{lccccccccc}
\hline\hline \addlinespace[0.15cm]
& \multicolumn{3}{c}{Strates 1 and 3} & \multicolumn{3}{c}{Strate 1}& \multicolumn{3}{c}{Strate 3}\\\cmidrule[0.2pt](l){2-4}\cmidrule[0.2pt](l){5-7}\cmidrule[0.2pt](l){8-10}
    & \multicolumn{1}{c}{obs.} & \multicolumn{1}{c}{mean}  & \multicolumn{1}{c}{st.dev.} & \multicolumn{1}{c}{obs.} & \multicolumn{1}{c}{mean}  & \multicolumn{1}{c}{st.dev.} & \multicolumn{1}{c}{obs.} & \multicolumn{1}{c}{mean}  & \multicolumn{1}{c}{st.dev.}\\\cmidrule[0.2pt](l){2-2}\cmidrule[0.2pt](l){3-3} \cmidrule[0.2pt](l){4-4} \cmidrule[0.2pt](l){5-5} \cmidrule[0.2pt](l){6-6}  \cmidrule[0.2pt](l){7-7}\cmidrule[0.2pt](l){8-8}\cmidrule[0.2pt](l){9-9}\cmidrule[0.2pt](l){10-10}
        & (1)& (2)& (3)& (4)& (5)& (6)& (7)& (8)& (9)\\   \addlinespace[0.1cm]\hline\addlinespace[0.10cm]
      &\multicolumn{9}{c}{2014}\\\cmidrule[0.2pt](l){2-10}\addlinespace[0.15cm]
                                      \multicolumn{1}{l}{\emph{\underline{Panel A}: Academic Outcomes (PASEC) }}            \\\addlinespace[0.15cm]
Reading score&        2088&        53.1&         9.2&        1578&        51.3&         8.8&         510&        58.6&         8.4\\
Reading score (females)&        1058&        53.6&         9.2&         802&        51.8&         8.5&         256&        59.4&         8.9\\
Reading score (males)&        1030&        52.6&         9.2&         776&        50.9&         9.0&         254&        57.7&         7.7\\
Math score  &        2088&        50.4&         8.7&        1578&        48.5&         8.0&         510&        56.2&         8.0\\
Math score (females)&        1058&        50.6&         8.7&         802&        48.6&         7.8&         256&        56.9&         8.5\\
Math score (males)&        1030&        50.1&         8.6&         776&        48.4&         8.2&         254&        55.5&         7.4\\

\addlinespace[0.3cm]
                   \multicolumn{1}{l}{\emph{\underline{Panel B}: Armed Conflict (ACLED) }}               \\\addlinespace[0.15cm]
Events involving any group&        2088&        34.4&        44.7&        1578&         9.0&         0.0&         510&       113.0&         0.0\\
Events involving Amazonian rebels&        2088&         0.0&         0.0&        1578&         0.0&         0.0&         510&         0.0&         0.0\\
Events involving Cameroonian army&        2088&         4.6&         8.2&        1578&         0.0&         0.0&         510&        19.0&         0.0\\
Fatalities by any group&        2088&       344.4&       569.0&        1578&        21.0&         0.0&         510&      1345.0&         0.0\\
Fatalities by Amazonian rebels&        2088&         0.0&         0.0&        1578&         0.0&         0.0&         510&         0.0&         0.0\\
Fatalities by Cameroonian army&        2088&        43.2&        76.1&        1578&         0.0&         0.0&         510&       177.0&         0.0\\

\addlinespace[0.15cm]\hline\addlinespace[0.15cm]
           \multicolumn{1}{l}{\emph{\underline{Panel C}: Academic Outcomes (PASEC)}}               &\multicolumn{9}{c}{2019}\\\cmidrule[0.2pt](l){2-10}\addlinespace[0.15cm]
 \addlinespace[0.15cm]
 Reading score&        2497&        55.9&         9.5&         727&        51.9&         9.4&        1770&        57.6&         9.1\\
Reading score (females)&        1279&        56.4&         9.6&         364&        52.3&         9.3&         915&        58.0&         9.2\\
Reading score (males)&        1218&        55.5&         9.5&         363&        51.5&         9.5&         855&        57.2&         9.0\\
Math score  &        2497&        52.4&         8.3&         727&        48.9&         7.5&        1770&        53.8&         8.1\\
Math score (females)&        1279&        52.2&         8.3&         364&        48.4&         7.2&         915&        53.8&         8.2\\
Math score (males)&        1218&        52.5&         8.2&         363&        49.5&         7.8&         855&        53.8&         8.1\\

 \addlinespace[0.3cm]
                   \multicolumn{1}{l}{\emph{\underline{Panel D}: Armed Conflict (ACLED) }}               \\\addlinespace[0.15cm]
Events involving any group&        2497&       435.9&        12.3&         727&       455.0&         0.0&        1770&       428.0&         0.0\\
Events involving Amazonian rebels&        2497&        47.8&        66.8&         727&       152.0&         0.0&        1770&         5.0&         0.0\\
Events involving Cameroonian army&        2497&       104.4&        89.5&         727&       244.0&         0.0&        1770&        47.0&         0.0\\
Fatalities by any group&        2497&       591.1&       131.3&         727&       796.0&         0.0&        1770&       507.0&         0.0\\
Fatalities by Amazonian rebels&        2497&        45.1&        70.4&         727&       155.0&         0.0&        1770&         0.0&         0.0\\
Fatalities by Cameroonian army&        2497&       226.0&       248.1&         727&       613.0&         0.0&        1770&        67.0&         0.0\\

 \addlinespace[0.15cm]
\hline\hline
\multicolumn{10}{p{15.6cm}}{\scriptsize{\textbf{Notes:} The sample in all columns is restricted to the period 2014 to 2019 and to pupils in the Anglophone subsystem, from pubic schools.  The sample in columns (1)-(3) includes strates 1 and 3 (see Table \ref{tab:stratesmain} for the definition of these strates). The sample in columns (4)-(6) is limited to the strate 1 (Zone Anglophone, public schools) and sample in columns (7)-(9) is limited to the strate 2 (Zone Francophone). The data on education outcomes comes from PASEC. The data on conflict comes from the ACLED.} }
\end{tabular}
\end{center}
\end{table}

To estimate the impact on pupil learning of Cameroon's Anglophone conflict, we integrate the PASEC data with data on conflict from the Armed Conflict Location \& Event Data Project (ACLED). The ACLED is a publicly available database that offers extensive information on armed conflicts, political violence, and protest events worldwide. It is a collaborative project between researchers, analysts, and organizations working in the fields of political violence, conflict resolution, and human rights.\footnote{The ACLED data can be accessed publicly at  \url{https://acleddata.com}}

ACLED  gathers and examines real-time information regarding the location, participants, casualties, and other pertinent details of conflict incidents. Their research encompasses various types of political violence, including state-based conflict, non-state conflict, and one-sided violence. The data undergoes regular updates, enabling the analysis of conflict trends and patterns over time. 

In our study, we utilize ACLED's data specifically pertaining to Cameroon, with a particular emphasis on the Anglophone conflict. Consequently, our scope is limited to violence associated with the actors involved in this conflict during the period from 2014 to 2019. The actors we identify as being associated with this conflict include the Ambazonian Separatists, the Ambazonian Defense Forces (ADF), and the armed and police forces of Cameroon.

As we describe in detail in Appendix \ref{Conflictdata}, in order to combine the ACLED data with the PASEC data, we construct indicators of conflict-related violence at the regional (or strate) level, however distinguishing between the armed groups involved. Descriptive statistics for our primary conflict violence variables are presented in Panels B and D of Table \ref{precharacteristics_tab}, organized by year.

\subsection{Distribution of Conflict Related Events}\label{Distribution_of_Conflict_Related_Events}

\begin{figure}[H]
             \caption{Conflict-Related Events and Fatalities Involving any Armed Group}
        \label{fig:ANY_conflict}
\begin{subfigure}{0.5\textwidth}
\caption{Events} \label{fig:Envent_conflict}
\includegraphics[width=\linewidth]{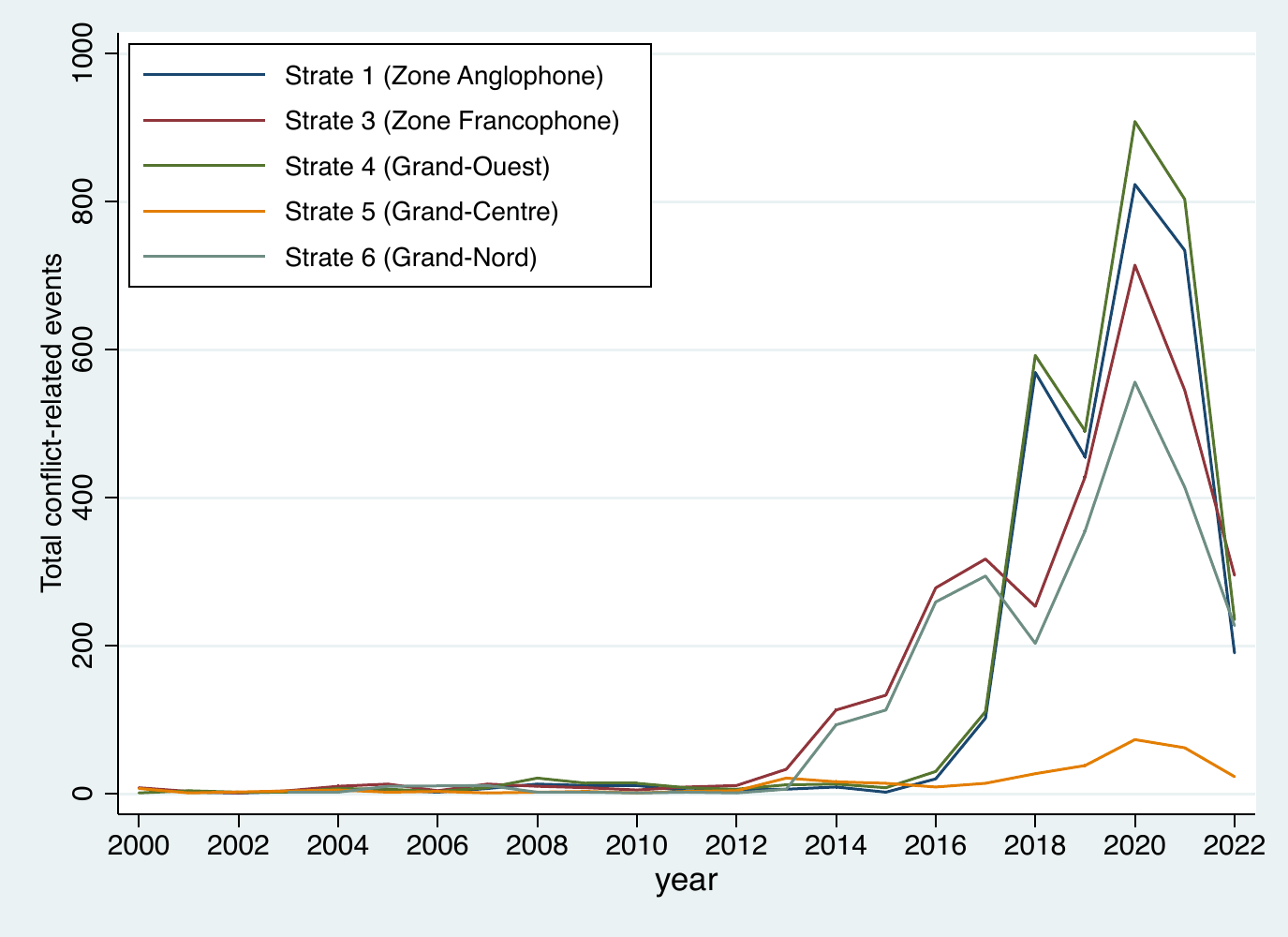}
\end{subfigure}\hspace*{\fill}
\begin{subfigure}{0.5\textwidth}
\caption{Fatalities} \label{fig:Fatal_conflict}
\includegraphics[width=\linewidth]{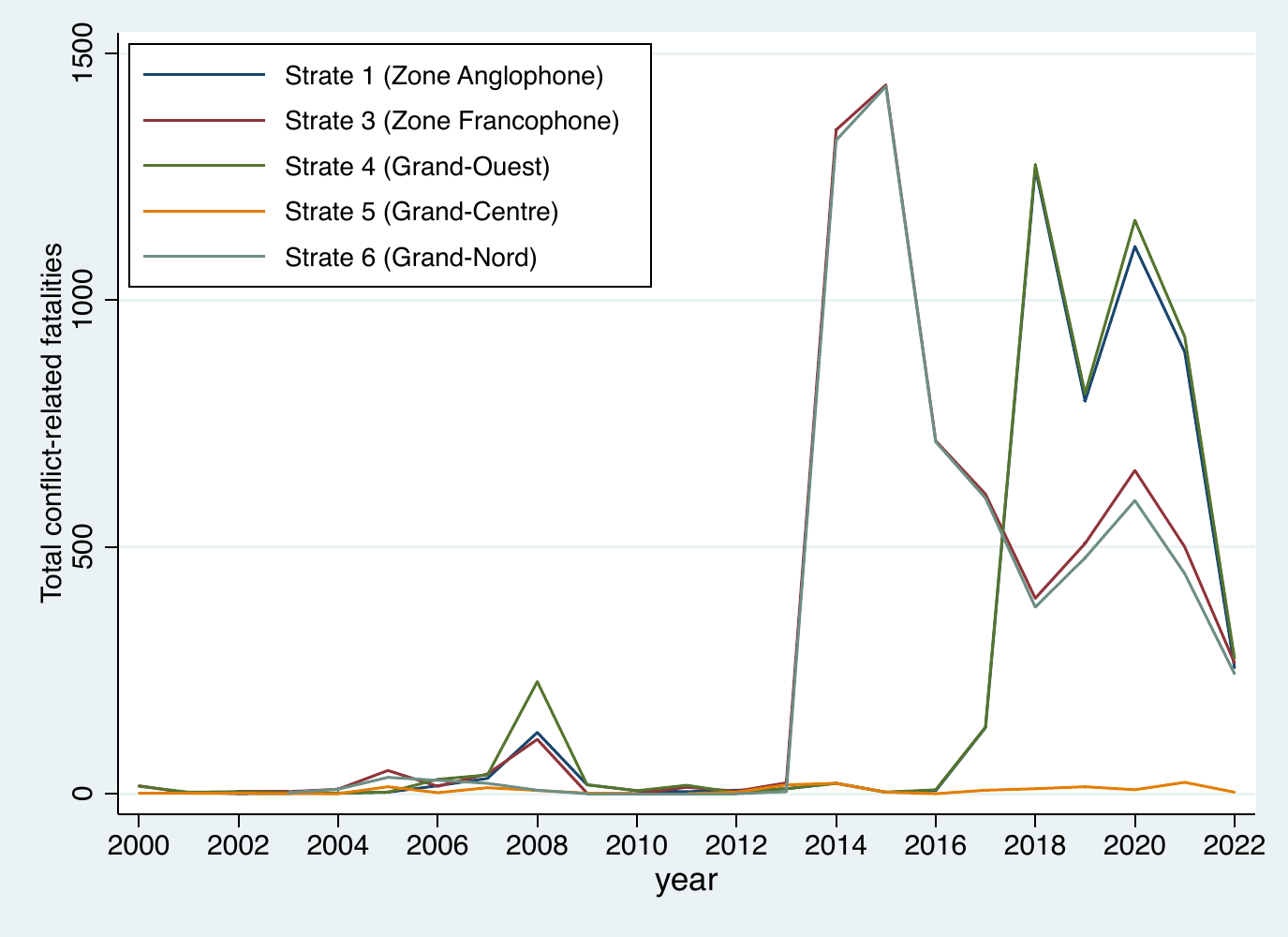}
\end{subfigure}
\end{figure}

Utilizing data extracted from the ACLED, Figure \ref{fig:Envent_conflict} provides an overview of the temporal distribution of all conflict-related incidents across the entirety of Cameroon's territory over the past two decades. As mentioned earlier, the majority of these incidents occurred after 2014. Complementing this, Figure \ref{fig:Fatal_conflict} specifically focuses on the number of fatalities resulting from these events, revealing significant regional variation between 2000 and 2022. Notably, Strates 1 and 4, identified as the epicenter of the Anglophone conflict, experienced the highest number of fatalities between 2014 and 2022.

\begin{figure}[H]
             \caption{Conflict-Related Events and Fatalities Involving Ambazonian Separatists}
        \label{fig:AMB_conflict}
\begin{subfigure}{0.5\textwidth}
\caption{Events} \label{fig:Envent_conflict_AMB}
\includegraphics[width=\linewidth]{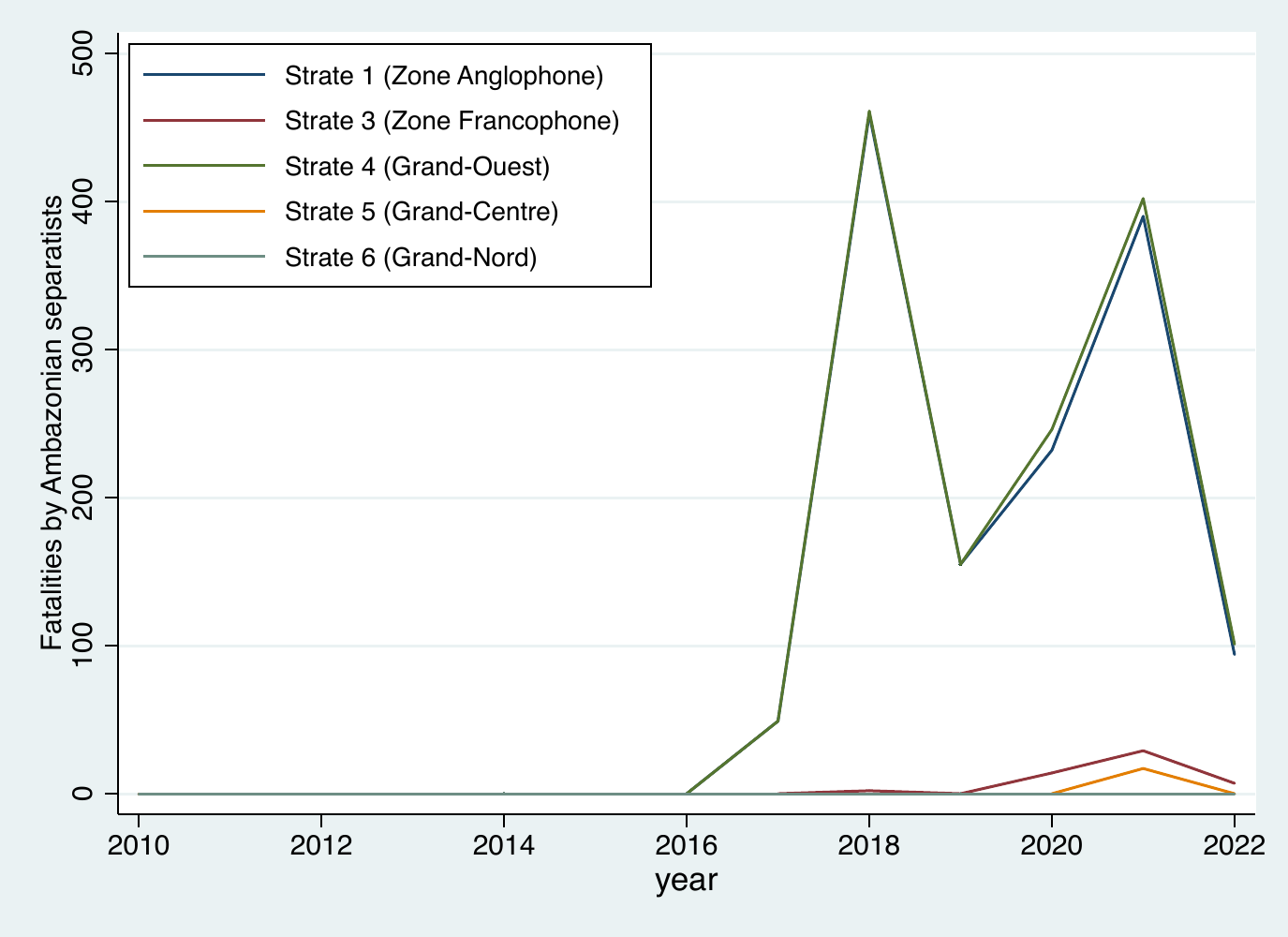}
\end{subfigure}\hspace*{\fill}
\begin{subfigure}{0.5\textwidth}
\caption{Fatalities} \label{fig:Envent_conflict_AMB}
\includegraphics[width=\linewidth]{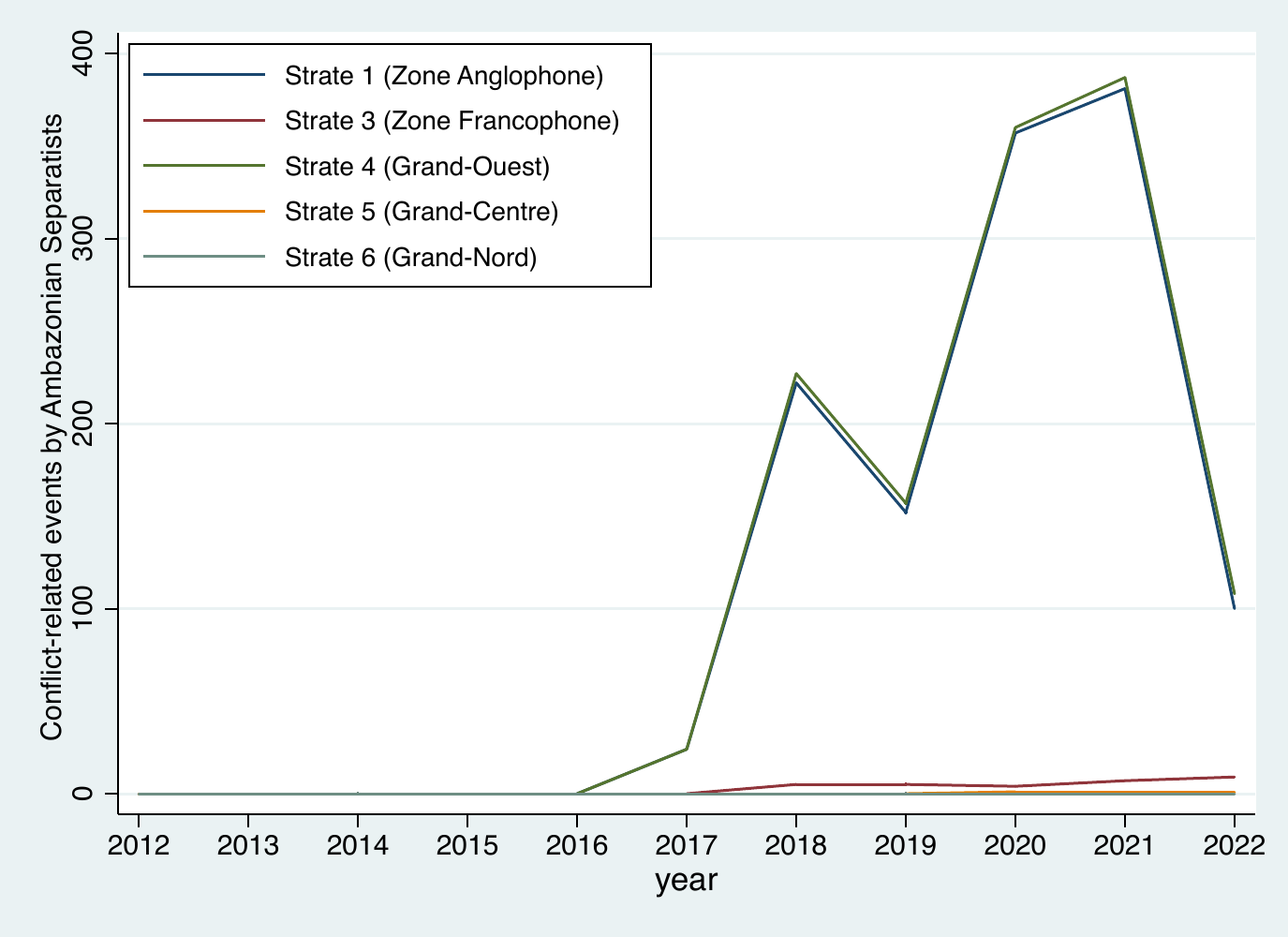}
\end{subfigure}
\end{figure}

Shifting our attention to events involving the Ambazonian Separatists, Figure \ref{fig:AMB_conflict} emphasizes the occurrence and intensity of such events exclusively within Strates 1 and 4 (North-West and South-West regions). Importantly, these events began escalating from 2016 onwards, and resulted in a significant number of fatalities between 2014 and 2022.

Figure \ref{fig:AMB_conflict}, when compared with Figure \ref{fig:Envent_conflict}, highlights a crucial fact: prior to 2016: while other conflict-related events occurred before 2016, they took place outside the strata of interest and involved armed actors unrelated to the Anglophone conflict (e.g., Boko Haram in the Far North region). This important finding, i.e. the localized nature of the Ambazonian separatists' actions and the timing of their occurrence, will play a pivotal role in our identification strategy.

% 
 %%%%%%%%%%%%%%%%%%%%%%%%%%%%%%%%%%%%%%%%%%%%%%%%%%%%%%%%%%%%%%%%%%%%%%%%%%%%

\section{Identification and Estimation Strategy} \label{Estimation_Strategy}

To assess the impact of violent conflict events during the Anglophone crisis on pupil learning in Cameroon, we propose a quasi-experimental approach that takes into account the level of exposure to conflict-related violence. Our analysis will specifically examine fatalities in the North-West and South-West regions, primarily involving actors such as the Ambazonian Separatists. Initially, our main focus will be on students enrolled in the Anglophone education subsystem. The presence of pupils in the anglophone subsystem both in the rest of the country and in the North-West and South-West regions allows for the design of a quasi-experiment. Subsequently, we will expand our analysis to include the Francophone education subsystem, which will provide supplementary results and will play a crucial role in evaluating our main identification assumption.

Specifically, for our main model, we define the following two groups:

\begin{itemize}
\item[(i)] Pupils who resided in regions where there was violence related to the Anglophone conflict and who were enrolled in the Anglophone education system. These pupils belong to strate 1.
\item[(ii)] Pupils who resided in regions unaffected by violence related to the Anglophone conflict and who also were enrolled in the Anglophone education system. These pupils belong to strate 3.
\end{itemize}

By comparing outcomes between two distinct groups - one exposed to conflict-related violence and the other unexposed - we can estimate the effect of violence on pupil learning using a quasi-experimental difference-in-differences approach.

More formally, let $Y_{ist}$ denote an outcome of interest, representing math or reading test scores for individual $i$ residing in region/stratum $r$ at time $t$. Here, $t$ can take values of 2014 or 2019, while $r$ can be either 1 or 3, corresponding to the two treatment groups. The index $i$ ranges from 1 to $N$.  To estimate the impact of the Anglophone conflict, we utilize the following model:
\begin{equation}\label{equa_1}
Y_{irt} =  a_0 +\alpha  AMB\_CE_{rt}+ c MIL\_CE_{rt}  + b X_{irt}  +a_1 dT+  u_{irt} 
\end{equation}            
where  $AMB\_CE_{rt}$ measures the intensity of conflict-related violent events involving Ambazonian separatists, such as the number of fatalities, in the region where individual  $i$ resided at time $t$. The dummy variable $dT$ indicates whether the data was collected in 2014 or 2019. $MIL\_CE_{rt}$ captures the intensity of conflict-related violent events involving the Cameroon army.\footnote{Fatalities involving the military are recorded in all regions and time periods.} The variable $X_{irt}$ represents a set of exogenous characteristics, including factors like sex, age, and grade, and $u_{irt}$ is the error term.

\medskip

To clearly establish the main identification assumptions of the model in equation (\ref{equa_1}), it is important to note that $AMB\_CE_{rt}$ is only observed in the North-West and South-West regions and only in 2019. Hence, we can rewrite equation (\ref{equa_1}) as follows:
\begin{equation}\label{equa_2}
Y_{irt} =  a_0 + a_1 dT +\alpha dT\times AMB\_CE_{r}+ c MIL\_CE_{rt} + b X_{irt} +d AMB\_CE_{r} +  u_{irt} 
\end{equation}       

In equation (\ref{equa_2}), the variable $AMB\_CE_{r}$ represents the intensity of conflict-related violent events in region $r$. It is important to note that for each individual $i$, region $r$, and time $t$, the expression $dT \times AMB\_CE_{r}$ in (\ref{equa_2}) is equivalent to $AMB\_CE_{rt}$ in (\ref{equa_1}). This equivalence implies that the model in (\ref{equa_1}) can be seen as a Difference-in-Difference style representation, which is essentially the same as the model (\ref{equa_2}).

Assuming that military-induced violence, including violent events and fatalities, has a consistent impact on the outcome variable $Y_{irt}$ across all regions, the parameter $\alpha$ captures the effect of the Anglophone conflict on $Y_{irt}$. Specifically, it measures the average effect of conflict violence, whether in terms of events or fatalities, on pupils enrolled in the Anglophone subsystem and residing in the North-West and South-West regions. This parameter represents the average treatment on the treated (ATT), providing an understanding of the average differential impact experienced by these pupils.

To evaluate this impact, we rely on the parallel trend assumption, which posits that in the absence of conflict-related events involving Ambazonian separatists, the trajectory of human capital accumulation would have been the same for pupils in the Anglophone system, regardless of their region of residence, between 2014 and 2019. This assumption is based on the lack of regional reforms in the educational system in Cameroon during that time period.

Moreover, we conduct  estimations to demonstrate that the violence associated with the conflict has negligible or no effect on pupils in the Francophone subsystem. These estimations serve as additional evidence, highlighting the specific influence of the Anglophone conflict on pupil outcomes and illustrating the divergent experiences between the Anglophone and Francophone subsystems.

In this study, we adopt a similar model to examine the impact of the conflict on human capital accumulation, explore potential transmission mechanisms, and conduct identifying assumption tests. Depending on the research question at hand, such as investigating transmission mechanisms, analyzing the effect of the conflict on human capital accumulation, or conducting identifying assumption tests, we adapt the outcome variable and regions while maintaining the core explanatory variables.

\section{Estimation of the Armed-Conflict Effects}

\subsection{Effects on Human Capital Accumulation}\label{maineffects}

In this section, we present the key findings regarding the impact of the Anglophone Armed conflict on early human capital accumulation. To capture the varying intensity of violence, we examine the estimated effects using two variables: the level of violence and the rate per 100,000 inhabitants.

The results presented in Table \ref{tab:conflict_full} specifically focus on students enrolled in the Anglophone education subsystem. These findings highlight that conflict-related violence in the Anglophone region significantly impedes the accumulation of human capital among students. Our analysis reveals that an additional ten violent events involving the Ambazonian separatists resulted in a 2.5\% decrease in language test scores and a 2.1\% decrease in mathematics test scores. Similarly, a ten-unit increase in conflict-related deaths led to a 1.9\% decline in language test scores and a 1.9\% reduction in mathematics test scores. Notably, these negative effects persist even when accounting for the region's population by measuring event occurrence and fatalities per 100,000 inhabitants.

Table \ref{tab:conflit_hetero} in the Appendix examines the heterogeneous effects based on the grade and gender of the students. We expect a potentially stronger effect on grade six pupils due to the conflict's duration covering a significant portion of their primary school education and the challenges they faced in a war-affected learning environment. In contrast, second-grade pupils experienced conflict throughout their entire primary schooling. However, the estimated effects on grade six pupils show a slightly more negative impact, although the difference is not statistically significant. Additionally, female pupils also exhibit a slightly more negative effect of violence on their human capital, although this difference is not statistically significant, at a 5\% level, either.

%%%%%%%%%%%%%%%%%%%%%%%%%%%%%

\begin{table}[H]
\begin{center}
{
\renewcommand{\arraystretch}{0.7}
\setlength{\tabcolsep}{10pt}
\caption {Effect of Conflict-related violence by Ambazonian Separatists on Academic Performance (period 2014-2019)}  \label{tab:conflict_full}
\small
\vspace{-0.3cm}\centering  \begin{tabular}{lcccc}
\hline\hline \addlinespace[0.15cm]
    \addlinespace[0.15cm]
    & (1)& (2)& (3)& (4)\\   \addlinespace[0.15cm]\hline\addlinespace[0.15cm]
           \multicolumn{1}{l}{\emph{\underline{Panel A}: }}               &\multicolumn{4}{c}{Dependent variable: reading score}\\\cmidrule[0.2pt](l){2-5}\addlinespace[0.15cm]
    &\multicolumn{1}{c}{Number}&\multicolumn{1}{c}{Rate}&\multicolumn{1}{c}{Number}&\multicolumn{1}{c}{Rate}\\\cmidrule[0.2pt](l){2-2} \cmidrule[0.2pt](l){3-4}\cmidrule[0.2pt](l){4-4}\cmidrule[0.2pt](l){5-5}\addlinespace[0.15cm]
    Events involving Ambazonian rebels&      -0.259\sym{***}&     -48.018\sym{***}&                     &                     \\
                    &     (0.042)         &     (8.124)         &                     &                     \\
Fatalities by Ambazonian rebels&                     &                     &      -0.193\sym{***}&     -35.025\sym{***}\\
                    &                     &                     &     (0.031)         &     (6.015)         \\
R-sq                &       0.111         &       0.111         &       0.100         &       0.100         \\
Observations        &        4585         &        4585         &        4585         &        4585         \\

    \addlinespace[0.15cm]\hline\addlinespace[0.15cm]
           \multicolumn{1}{l}{\emph{\underline{Panel B}: }}               &\multicolumn{4}{c}{Dependent variable: math score}\\\cmidrule[0.2pt](l){2-5}\addlinespace[0.15cm]
                          &\multicolumn{1}{c}{Number}&\multicolumn{1}{c}{Rate}&\multicolumn{1}{c}{Number}&\multicolumn{1}{c}{Rate}\\\cmidrule[0.2pt](l){2-2} \cmidrule[0.2pt](l){3-4}\cmidrule[0.2pt](l){4-4}\cmidrule[0.2pt](l){5-5}\addlinespace[0.15cm]
                        Events involving Ambazonian rebels&      -0.211\sym{***}&     -38.420\sym{***}&                     &                     \\
                    &     (0.038)         &     (7.284)         &                     &                     \\
Fatalities by Ambazonian rebels&                     &                     &      -0.191\sym{***}&     -34.753\sym{***}\\
                    &                     &                     &     (0.029)         &     (5.643)         \\
R-sq                &       0.099         &       0.099         &       0.121         &       0.121         \\
Observations        &        4585         &        4585         &        4585         &        4585         \\

\addlinespace[0.15cm]\hline\hline\addlinespace[0.15cm]
\multicolumn{5}{p{14.5cm}}{\footnotesizes{\textbf{Notes:} All columns in report the estimates from Eq. (\ref{equa_1}).  Rate is per 100 000 inhabitants. Robust standard errors (in parentheses) are clustered by school.  Do not include fixed effects by school. * denotes statistically significant estimates at 10\%, ** denotes significant at 5\% and *** denotes significant at 1\%.} }
\end{tabular}
}
\end{center}
\end{table}

%%%%%%%%%%%%%%%%%%%%%%%%%%%%%%

\subsection{Discussion of Transmission Mechanisms}

The findings presented in the previous section reveal a detrimental impact of violent events stemming from the Anglophone conflict on the early accumulation of human capital. In this section, we examine potential mechanisms through which the acquisition of language and mathematics skills by pupils is hindered, shedding light on their diminished learning outcomes. We explore and discuss two distinct yet interconnected transmission channels.

Firstly, we analyze the effect of conflict-related fatalities on teachers' absenteeism, as illustrated in Table \ref{tab:mechanism} (columns 1 and 2). These estimates indicate that an increase in fatalities resulting from events involving Ambazonian separatists leads to a higher probability of teacher absenteeism. This decrease in teachers' presence is observed in both grades 2 and 6. We contend that, from the perspective of educators and adults, the heightened risk associated with the conflict detrimentally affects the quality and quantity of pupils' learning experiences, thereby impeding their skill acquisition. For instance, the presence of conflict may trigger temporary migration to safer areas among both pupils and teachers, resulting in learning gaps.

Secondly, we posit that the presence of conflict contributes to economic hardship, potentially diverting pupils from their studies to engage in work within the legal or illegal economy, further exacerbating learning gaps. Table \ref{tab:mechanism} (column 3) provides estimates that shed light on this relationship. They reveal that an increase in conflict-related fatalities in events involving Ambazonian separatists decreases the likelihood of a school having access to electricity. This suggests that the violence associated with the conflict may have led to the destruction of existing infrastructure or created economic challenges that prioritize other urgent needs over educational resources.

\begin{table}[H]
\begin{center}
{
\renewcommand{\arraystretch}{0.7}
\setlength{\tabcolsep}{10pt}
\caption {Effect of Conflict-related  violence by Ambazonian Separatists on  teacher absenteeism and access to electricity}  \label{tab:mechanism}
\small
\vspace{-0.3cm}\centering  \begin{tabular}{lcccc}
\hline\hline \addlinespace[0.15cm]
    \addlinespace[0.15cm]
    & (1)& (2) & (3) \\   \addlinespace[0.15cm]\hline\addlinespace[0.15cm]
     &\multicolumn{3}{c}{Dependents variables:}\\\addlinespace[0.15cm]
    &\multicolumn{1}{c}{Teacher}&\multicolumn{1}{c}{Teacher}&\multicolumn{1}{c}{Electricity}\\
    &\multicolumn{1}{c}{absent in Grade 2}&\multicolumn{1}{c}{absent in Grade 6}&\multicolumn{1}{c}{in school}\\\cmidrule[0.2pt](l){2-2} \cmidrule[0.2pt](l){3-3}\cmidrule[0.2pt](l){4-4}\addlinespace[0.15cm]
   Fatalities by Ambazonian rebels (rate)&       2.037\sym{***}&       1.288\sym{***}&      -1.222\sym{***}\\
                    &     (0.513)         &     (0.451)         &     (0.213)         \\
R-sq                &       0.118         &       0.042         &       0.051         \\
Observations        &        1802         &        3436         &        3150         \\

\addlinespace[0.15cm]\hline\hline\addlinespace[0.15cm]
\multicolumn{4}{p{15cm}}{\footnotesizes{\textbf{Notes:} All columns in report the estimates from Eq. (\ref{equa_1}).  Rate is per 10000 inhabitants. Robust standard errors (in parentheses) are clustered by school.  * denotes statistically significant estimates at 10\%, ** denotes significant at 5\% and *** denotes significant at 1\%.} }
\end{tabular}
}
\end{center}
\end{table}
 
 %%%%%%%%%%%%%%%%%%%%%%%%%%%%%%%%%%%%%%%%%%%%%%%%%%%%%%%%%%%%%%%%%%%%%%
%\section{Conflict effects on the Francophone Subsystem} 
%
%\subsection{Effects Estimation}

\section{Identifying Assumptions Checks: Effects on the Francophone Subsystem}

In this section, we analyze the impact of the Anglophone conflict on Francophone pupils, focusing on two key objectives. Firstly, we assess the implications arising from the existence of two separate educational systems in the context of this conflict. By closely examining the experiences of Francophone pupils, we aim to shed light on any disparities or divergences that may emerge as a result of this unique socio-political situation. Secondly, we use this analysis to validate the plausibility of our identifying assumptions. 

In our research design, outlined in Section \ref{Estimation_Strategy}, we primarily studied a sample of pupils within the Anglophone subsystem. Now, our attention turns to evaluating the influence of the Anglophone conflict on pupils belonging to the Francophone subsystem. To achieve this, we designate the Grand-Ouest (4) and Grand-Nord (6) strata as the treatment and control groups, respectively. 

The estimation results, presented in Table \ref{tab:placebo}, yield significant insights into the effect of conflict-related violence on human capital accumulation within the Francophone subsystem. In particular, we find that an increase in violent events or deaths does not have a significant impact on the level of human capital accumulated in both language and mathematics. Specifically, we observe a negligible decrease in mathematics skill acquisition and a slight increase in language skill acquisition. However, these effects are minimal compared to those observed in the sample of Anglophone pupils residing in conflict-affected regions (North-West and South-West). Thus,  pupils in the Francophone educational subsystem were largely unaffected. Indeed, the effects of conflicts are approximately 10 times smaller than the effects observed in the Anglophone educational subsystem sample.

\begin{table}[H]
\begin{center}
{\renewcommand{\arraystretch}{0.7}
\setlength{\tabcolsep}{10pt}
\caption {Effect of Conflict-related violence by Ambazonian Separatists on Academic Performance: Placebo tests (Francophone system in regions Grand Ouest vs Grand Nord)}  \label{tab:placebo}
\small
\vspace{-0.3cm}\centering  \begin{tabular}{lcccc}
\hline\hline \addlinespace[0.15cm]
    \addlinespace[0.15cm]
    & (1)& (2)& (3)& (4)\\   \addlinespace[0.15cm]\hline\addlinespace[0.15cm]
           \multicolumn{1}{l}{\emph{\underline{Panel A}: }}               &\multicolumn{4}{c}{Dependent variable: reading score}\\\cmidrule[0.2pt](l){2-5}\addlinespace[0.15cm]
    &\multicolumn{1}{c}{Number}&\multicolumn{1}{c}{Rate}&\multicolumn{1}{c}{Number}&\multicolumn{1}{c}{Rate}\\\cmidrule[0.2pt](l){2-2} \cmidrule[0.2pt](l){3-4}\cmidrule[0.2pt](l){4-4}\cmidrule[0.2pt](l){5-5}\addlinespace[0.15cm]
    Events involving Ambazonian rebels&       0.014\sym{*}  &       1.219\sym{*}  &                     &                     \\
                    &     (0.008)         &     (0.715)         &                     &                     \\
Fatalities by Ambazonian rebels&                     &                     &       0.014\sym{*}  &       1.235\sym{*}  \\
                    &                     &                     &     (0.008)         &     (0.724)         \\
R-sq                &       0.015         &       0.015         &       0.015         &       0.015         \\
Observations        &        4437         &        4437         &        4437         &        4437         \\

    \addlinespace[0.15cm]\hline\addlinespace[0.15cm]
           \multicolumn{1}{l}{\emph{\underline{Panel B}: }}               &\multicolumn{4}{c}{Dependent variable: math score}\\\cmidrule[0.2pt](l){2-5}\addlinespace[0.15cm]
                          &\multicolumn{1}{c}{Number}&\multicolumn{1}{c}{Rate}&\multicolumn{1}{c}{Number}&\multicolumn{1}{c}{Rate}\\\cmidrule[0.2pt](l){2-2} \cmidrule[0.2pt](l){3-4}\cmidrule[0.2pt](l){4-4}\cmidrule[0.2pt](l){5-5}\addlinespace[0.15cm]
                        Events involving Ambazonian rebels&      -0.019\sym{**} &      -0.703\sym{**} &                     &                     \\
                    &     (0.008)         &     (0.270)         &                     &                     \\
Fatalities by Ambazonian rebels&                     &                     &      -0.020\sym{***}&      -0.694\sym{***}\\
                    &                     &                     &     (0.008)         &     (0.264)         \\
R-sq                &       0.033         &       0.035         &       0.035         &       0.035         \\
Observations        &        4585         &        4585         &        4585         &        4585         \\

\addlinespace[0.15cm]\hline\hline\addlinespace[0.15cm]
\multicolumn{5}{p{14.5cm}}{\footnotesizes{\textbf{Notes:} All columns in report the estimates from Eq. (\ref{equa_1}).  Rate is per 100000 inhabitants. Robust standard errors (in parentheses) are clustered by school.  No model includes fixed effects by school. * denotes statistically significant estimates at 10\%, ** denotes significant at 5\% and *** denotes significant at 1\%.} }
\end{tabular}
}
\end{center}
\end{table}

As mentioned earlier, the results in Table \ref{tab:placebo} not only provide new and relevant evidence on the connection between the Anglophone conflict and the educational outcomes of students in the Francophone education subsystem, but also support the causal interpretation of our estimated effects by aligning with our primary identification assumption: the parallel trend assumption.

Specifically, Table \ref{tab:placebo} demonstrates that when comparing educational outcomes between 2014 and 2019, within two groups that closely resemble those examined in our baseline specification, but with significantly lower exposure to the Anglophone conflict (as anticipated for students residing in the same regions as those included in the baseline specification, yet enrolled in the Francophone educational subsystem which was not directly affected by the Anglophone conflict due to its concentrated impact on Anglophone schools),\footnote{It is worth noting that while some pupils from the Grand-Ouest strata may originate from conflict-affected regions, such as the North-West and South-West, the majority of them come from the West and Littoral regions.} there is no substantial effect of the Anglophone conflict on the accumulation of human capital. 

\medskip

To summarize, this section's results make two important contributions. Firstly, they provide crucial evidence that the observed disparities found in Section \ref{maineffects} can be attributed to the Anglophone conflict rather than other confounding factors. Secondly, these findings shed light on how linguistic disparities are further exacerbated within the context of armed conflict. Specifically, they demonstrate that despite the Anglophone conflict originating from the existence of such disparities and the aim to reduce them, the conflict paradoxically leads to an increase in these disparities. This increase is evident not only in the short term, where there was direct loss of lives and destruction of infrastructure, particularly in English-speaking regions, but also in the long term, affecting the academic outcomes of a generation of Anglophone students, which significantly lag behind those of their Francophone counterparts, and it is likely that these negative early-life experiences play a crucial role in shaping individuals' future educational and economic trajectories.

\section{Conclusion}

This paper aims to quantify the consequences of conflict-related violence in the Anglophone crisis on human capital accumulation, with a specific focus on the educational outcomes of Cameroonian Anglophone pupils. By utilizing high-quality individual-level data on pupils in Cameroon collected by the PASEC from 2014 to 2019, combined with precise information on conflict-related violent events from the ACLED, we present a causal estimate of the Anglophone conflict's effects.

Employing a difference-in-difference design, our findings reveal that an additional hundred violent events linked to the Ambazonian rebels result in a 2.5\% reduction in language test scores and a 2.1\% decrease in mathematics test scores. Likewise, a hundred more conflict-related deaths lead to a 1.9\% decline in language test scores and a 1.9\% reduction in mathematics test scores. The burden of the conflict is disproportionately borne by pupils within the Anglophone educational subsystem. Empirical evidence further suggests that conflict violence increases absenteeism among both teachers and pupils, thereby diminishing the quality and quantity of learning interactions and exacerbating the learning gap. Additionally, we provide evidence of heightened economic hardship stemming from conflict-related violence.

The short-term impact of the Anglophone conflict on human capital accumulation is substantial, and the consequences could extend beyond the immediate learning gap, given the critical role of early human capital in adults' health, economic, and social well-being. The evidence presented in this paper offers valuable insights for post-conflict reconstruction efforts in the realm of human capital capacity building in the Anglophone conflict in Cameroon.

\newpage

\bibliographystyle{econometrica}

\bibliography{conflict_Educ.bib}

%%%%%%%%%%%%%%%%%%%%%%%%%%%%%%%%%%%%%%%%%%%%%%%%%%%%%%%%%%%%%%%%
%%%%%%%%%%%%%%%%%%%%%%%%%%%%%%%%%%%%%%%%%%%%%%%%%%%%%%%%%%%%%%%%
%%%%%%%%%%%%%%%%%%%%%%%%%%%%%%%%%%%%%%%%%%%%%%%%%%%%%%%%%%%%%%%%
%%%%%%%%%%%%%%%%%%%%%%%%%%%%%%%%%%%%%%%%%%%%%%%%%%%%%%%%%%%%%%%%

\hbox {} \newpage

\appendix

\section{Appendix}

\setcounter{table}{0}
\setcounter{figure}{0}
\renewcommand{\thefigure}{\Alph{section}\arabic{figure}}
\renewcommand{\thetable}{\Alph{section}\arabic{table}}

\subsection{Additional Figures and Tables}
\begin{figure}[H]
\centering
\caption{Fatalities by Ambazonian Rebels between 2014 and 2019}
\label{Fatal_AMB_map}
\resizebox{12cm}{16cm}{\includegraphics[width=0.8\linewidth]{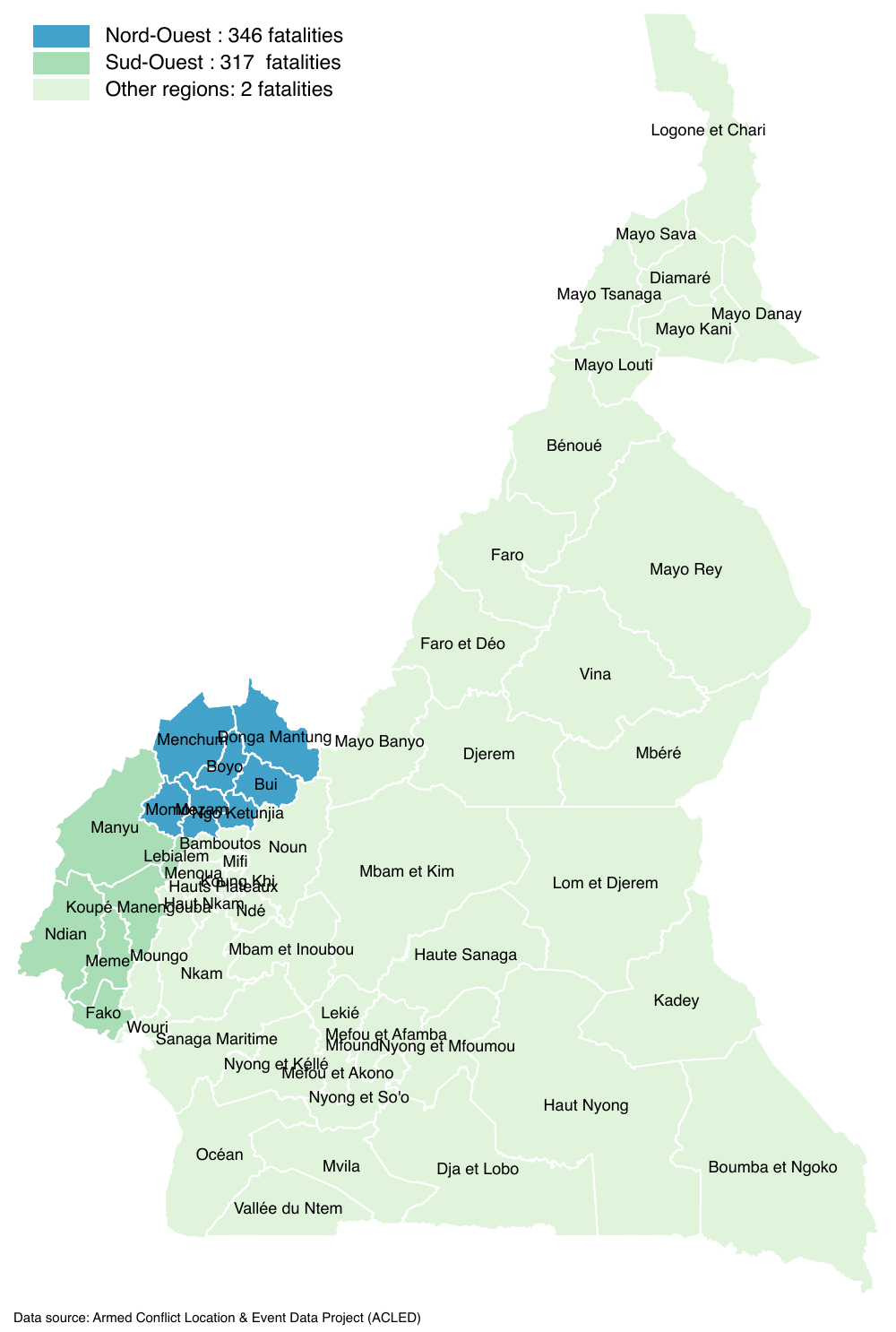}}
\end{figure}

%%%%%%%%%%%%%%%%%%%%%%%%%%%%%%

\newpage
\begin{table}[H]
\begin{center}
{
\renewcommand{\arraystretch}{0.7}
\setlength{\tabcolsep}{10pt}
\caption {Effect of Conflict-related  violence by Ambazonian Separatists on Academic Performance (period 2014-2019): Heterogeneous effects}  \label{tab:conflit_hetero}
\small
\vspace{-0.3cm}\centering  \begin{tabular}{lcccc}
\hline\hline \addlinespace[0.15cm]
    \addlinespace[0.15cm]
    & (1)& (2) & (3)& (4) \\   \addlinespace[0.15cm]\hline\addlinespace[0.15cm]
           \multicolumn{1}{l}{\emph{\underline{Panel A}: }}               &\multicolumn{4}{c}{Dependent variable: reading score}\\\cmidrule[0.2pt](l){2-5}\addlinespace[0.15cm]
    &\multicolumn{1}{c}{Number}&\multicolumn{1}{c}{Rate}&\multicolumn{1}{c}{Number}&\multicolumn{1}{c}{Rate}\\\cmidrule[0.2pt](l){2-2} \cmidrule[0.2pt](l){3-3}\cmidrule[0.2pt](l){4-4}  \cmidrule[0.2pt](l){5-5}\addlinespace[0.15cm]
    Fatalities by Ambazonian rebels&      -0.197\sym{***}&     -31.799\sym{***}&      -0.184\sym{***}&     -33.316\sym{***}\\
                    &     (0.048)         &     (9.168)         &     (0.031)         &     (5.994)         \\
Fatalities by Ambazonian rebels $\times$ grade 6&       0.008         &      -3.940         &                     &                     \\
                    &     (0.046)         &     (8.820)         &                     &                     \\
Fatalities by Ambazonian rebels $\times$ female&                     &                     &      -0.015         &      -3.394         \\
                    &                     &                     &     (0.018)         &     (3.233)         \\
R-sq                &       0.109         &       0.109         &       0.087         &       0.101         \\
Observations        &        4585         &        4585         &        4585         &        4585         \\

    \addlinespace[0.15cm]\hline\addlinespace[0.15cm]
           \multicolumn{1}{l}{\emph{\underline{Panel B}: }}               &\multicolumn{4}{c}{Dependent variable: math score}\\\cmidrule[0.2pt](l){2-5}\addlinespace[0.15cm]
    &\multicolumn{1}{c}{Number}&\multicolumn{1}{c}{Rate}&\multicolumn{1}{c}{Number}&\multicolumn{1}{c}{Rate}\\\cmidrule[0.2pt](l){2-2} \cmidrule[0.2pt](l){3-3}\cmidrule[0.2pt](l){4-4}  \cmidrule[0.2pt](l){5-5}\addlinespace[0.15cm]
                       Fatalities by Ambazonian rebels&      -0.178\sym{***}&     -29.731\sym{***}&      -0.178\sym{***}&     -32.825\sym{***}\\
                    &     (0.044)         &     (8.496)         &     (0.029)         &     (5.662)         \\
Fatalities by Ambazonian rebels $\times$ grade 6&      -0.015         &      -6.351         &                     &                     \\
                    &     (0.044)         &     (8.468)         &                     &                     \\
Fatalities by Ambazonian rebels $\times$ female&                     &                     &      -0.029\sym{*}  &      -3.834         \\
                    &                     &                     &     (0.016)         &     (3.037)         \\
R-sq                &       0.125         &       0.125         &       0.106         &       0.121         \\
Observations        &        4585         &        4585         &        4585         &        4585         \\

\addlinespace[0.15cm]\hline\hline\addlinespace[0.15cm]
\multicolumn{5}{p{16.2cm}}{\footnotesizes{\textbf{Notes:} All columns in report the estimates from Eq. (\ref{equa_1}).  Rate is per 100000 inhabitants. Robust standard errors (in parentheses) are clustered by school.  Do not include fixed effects by school. * denotes statistically significant estimates at 10\%, ** denotes significant at 5\% and *** denotes significant at 1\%.} }
\end{tabular}
}
\end{center}
\end{table}

\subsection{Education and Conflict Data}

 %%%%%%%%%%%%%%%%%%%%%%%%%%%%%%%%%%%%%%%%%%%%%%%%%
\subsubsection{Data on educational outcomes}\label{Educationdata}

The data used in this paper to examine educational outcomes is obtained from the  Programme d'analyse des Syst\`emes \'educatifs de la Confemen (PASEC). This dataset encompasses information for two distinct time periods, specifically 2014 and 2019, and concentrates on two specific grade levels, namely grade 2 and grade 6.  It encompasses data concerning the regional geographical location of students and their enrollment in either the francophone or anglophone educational subsystem.  In this study, we use the term `strate' to refer to the combination of a student's region and their educational subsystem.

Importantly, the information concerning these "strates" differs between the two available years of data. In 2019, PASEC classifies students into 12 distinct strates, as depicted in column 1 of Table \ref{strates1419}. Conversely, for the year 2014, the classification comprises only 6 strates, as shown in column 2 of Table \ref{strates1419}. 

\begin{table}[H]
\centering 
{\renewcommand{\arraystretch}{1}
\setlength{\tabcolsep}{10pt}
\caption{Strates in  2014  and 2019}
\small
  \begin{tabular}{ll}
\hline\hline \addlinespace[0.1cm]
            \multicolumn{1}{c}{(1)} & \multicolumn{1}{c}{(2)} \\\addlinespace[0.1cm]\cmidrule[0.2pt](l){1-1}\cmidrule[0.2pt](l){2-2}
   \multicolumn{1}{c}{\textbf{Strates in 2019}} & \multicolumn{1}{c}{\textbf{Strates in 2014}}  \\\addlinespace[0.1cm]\hline\addlinespace[0.1cm]
   1 Ouest francophone  &  1 Zone anglophone publique\\
2 Littoral francophone & 2 Zone anglophone priv\'ee \\
3 Centre francophone & 3 Zone francophone  \\
4 Est francophone &  4 Grand-Ouest \\
5 Sud francophone  &  5 Grand-Centre\\
6 Adamaoua francophone  & 6 Grand-Nord \\
7 Extreme-Nord francophone   \\
8 Nord francophone \\
9 Ouest anglophone   \\
10 Centre anglophone  \\
11 Littoral anglophone  \\
12 Reste anglophone& \\
\addlinespace[0.1cm]\hline\hline\addlinespace[0.1cm]
\end{tabular}
\label{strates1419}
}
\end{table}

In our identification strategy we exploit regional, education subsystem and temporal variations. Therefore, understanding the process of matching `strates' between different years is crucial. It is important to note that the matching between `strates' across years is feasible because the `strates' defined in one year are always subsets of the `strates' defined in another year. This strict overlapping allows us to successfully match strata from one year to another.

Table \ref{strates1419equ} provides a comprehensive breakdown of the methodology employed for the matching process. Specifically, the table delineates the  mapping between different elements. The initial eight lines pertain to the francophone system, illustrating the precise geographic correspondence between each of the first eight strates in column (1) of Table \ref{strates1419} and a single strate in column (2). Similarly, lines 9 to 11 delineate the mapping for the anglophone system.

\begin{table}[H]
\centering 
{\renewcommand{\arraystretch}{1}
\setlength{\tabcolsep}{10pt}
\caption{Strate matching between strates in 2014 and 2019}
\small
  \begin{tabular}{lcl}
\hline\hline \addlinespace[0.15cm]
            \multicolumn{1}{c}{(1)} & & \multicolumn{1}{c}{(2)} \\\addlinespace[0.15cm]\cmidrule[0.2pt](l){1-1}\cmidrule[0.2pt](l){3-3}
   \multicolumn{1}{c}{\textbf{Strates in 2019 dataset}} & & \multicolumn{1}{c}{\textbf{Strates in 2014 dataset}}  \\\addlinespace[0.15cm]\hline\addlinespace[0.1cm]
   1 Ouest francophone  & $\rightarrow $ &  4 Grand-Ouest \\
2 Littoral francophone & $\rightarrow $ & 4 Grand-Ouest \\
3 Centre francophone & $\rightarrow $ & 5 Grand-Centre \\
4 Est francophone & $\rightarrow $ &  5 Grand-Centre \\
5 Sud francophone & $\rightarrow $  &  5 Grand-Centre \\
6 Adamaoua francophone & $\rightarrow $  & 6 Grand-Nord \\
7 Extreme-Nord francophone  & $\rightarrow $ & 6 Grand-Nord \\
8 Nord francophone & $\rightarrow $ & 6 Grand-Nord \\
9 Ouest anglophone  & $\rightarrow $ & 3 Zone francophone \\
10 Centre anglophone & $\rightarrow $ & 3 Zone francophone \\
11 Littoral anglophone & $\rightarrow $ & 3 Zone francophone \\
12 Reste anglophone & $\rightarrow $ & 1 Zone anglophone publique  \\ &&2 Zone anglophone priv\'ee \\
\addlinespace[0.1cm]\hline\hline
\end{tabular}
\label{strates1419equ}
}
\end{table}

\begin{figure}[H]
             \caption{Map of Cameroon with representation of the different zones (strates)}
        \label{fig:mapstrates2014}
        \vspace{-0.0cm}
\begin{subfigure}{0.5\textwidth}
\includegraphics[width=0.75\textwidth]{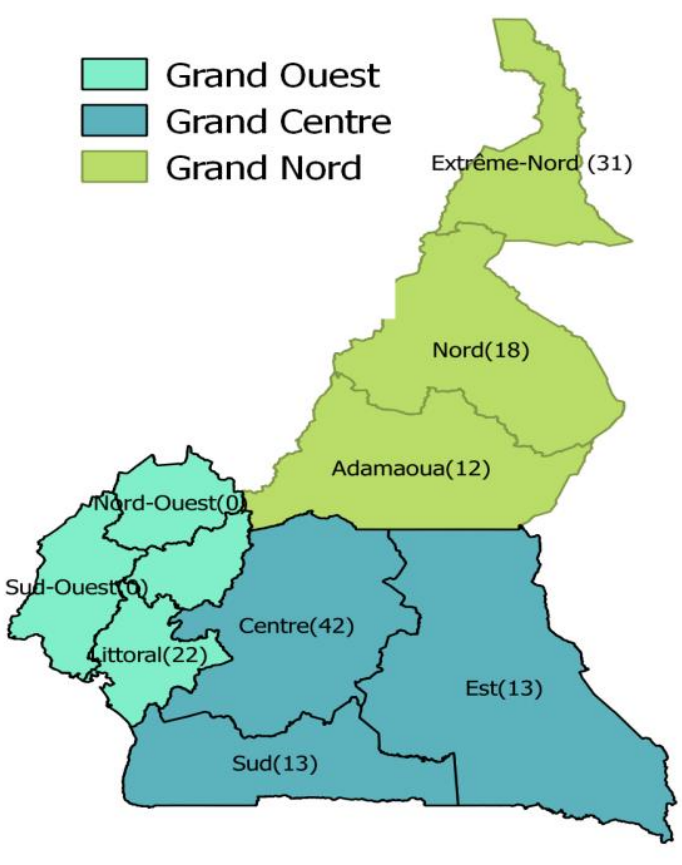}
\caption{Francophone subsystem} \label{fig:mapstrates2014A}
\end{subfigure}\hspace*{0.1cm}
\begin{subfigure}{0.5\textwidth}
\includegraphics[width=0.75\textwidth]{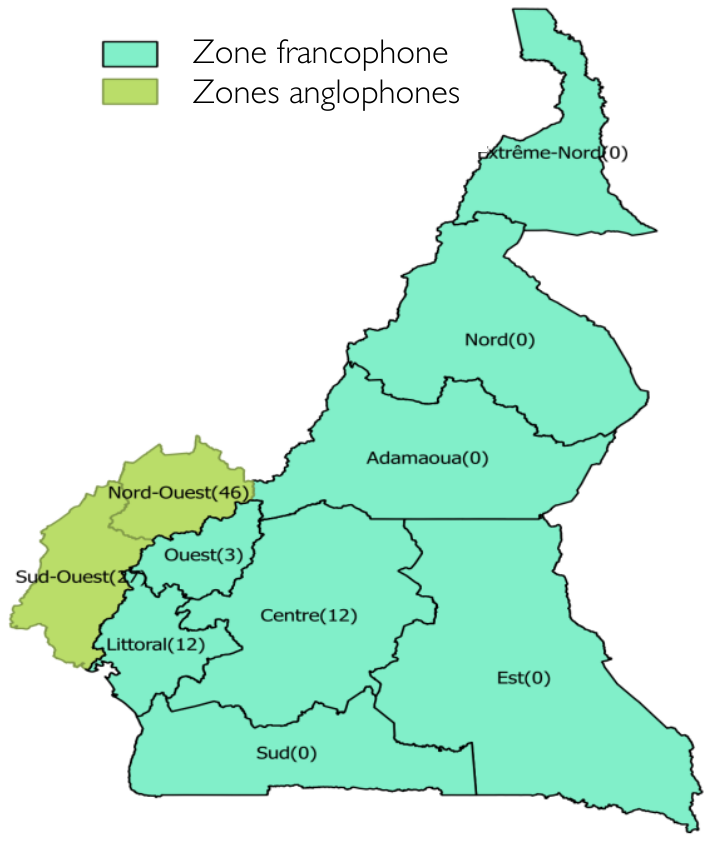}
\caption{Anglophone subsystem} \label{fig:mapstrates2014B}
\end{subfigure}
\end{figure}

The geographic correspondences, specific to an educational subsystem, are illustrated in Figure \ref{fig:mapstrates2014}. The left map (map \ref{fig:mapstrates2014A}) displays the matching for the Francophone subsystem, while the right map (map \ref{fig:mapstrates2014B}) represents the matching for the anglophone subsystem. Additionally, referring to line 12 in Table \ref{strates1419equ}, which represents two anglophone subregions in 2019, these subregions were classified as the 'Zone anglophone' (both public and private) in 2014. As the 2019 data does not differentiate between these two subregions, we associate them with the 'reste anglophone' strate in 2019.

 %%%%%%%%%%%%%%%%%%%%%%%%%%%%%%%%%%%%%%%%%%%%%%%%%
\subsubsection{Conflict data}\label{Conflictdata}

The paper relies on conflict data obtained from the Armed Conflict Location and Event Data Project (ACLED). ACLED offers comprehensive information on a wide range of violent and non-violent events involving political agents. Our analysis focuses on the period between 2014 and 2019. The conflict-related event data includes various aspects such as event type (e.g., battles, violence against civilians, explosions), date (day, month, year), location (geographic coordinates and municipality), and the armed actors involved (e.g., Ambazonian Separatists, Military Forces of Cameroon).

As the educational outcome data (from PASEC) only provides information at the `strate' and year levels (as discussed in Sub-section \ref{Educationdata}), we initially aggregated the conflict-related events indicators at these levels. For each year and strate, we tallied the number of events and fatalities by armed group and type of violence. This allowed us to create conflict-related indicators that could be matched with the educational outcomes data.

Next, we focused on identifying the specific armed actors directly linked to Cameroon's anglophone conflict. Our analysis distinguished between two types of armed groups: rebel groups and groups associated with state forces. The rebel groups identified were the Ambazonian Separatists and the Ambazonia Defense Forces (ADF). The groups associated with state forces were the Military Forces of Cameroon and the Police Forces of Cameroon. We then included the number of conflict events and fatalities within these groups, encompassing all types of violence. Importantly, the violence perpetrated by these two types of armed groups accounted for 100\% of the conflict-related events that took place within the regions (or strates) concentrated in our study, as stated in the main text.

\end{document}